\documentclass[a4paper,11pt]{article}
\usepackage{bm}
\usepackage{pos}
\usepackage{subcaption}
\DeclareMathOperator{\tr}{tr}
\DeclareMathOperator{\re}{Re}

\title{New gauge-independent transition separating confinement-Higgs phase in the lattice gauge-fundamental scalar model}

\author*[a]{Ryu Ikeda}
\author[b]{Kei-Ichi Kondo}
\author[c,d]{Akihiro Shibata}
\author[e]{Seikou Kato}

\affiliation[a]{Department of Physics, Graduate School of Science and Engineering, Chiba University, Chiba 263-8522, Japan}
\affiliation[b]{Department of Physics, Graduate School of Science, Chiba University, Chiba 263-8522, Japan}
\affiliation[c]{Computing Research Center, High Energy Accelerator Research Organization (KEK), Tsukuba 305-0801, Japan}
\affiliation[d]{Department of Accelerator Science, SOKENDAI (The Graduate University for Advanced Studies), Tsukuba 305-0801, Japan}
\affiliation[e]{Oyama National College of Technology, Oyama 323-0806, Japan}

\emailAdd{cdna0955@chiba-u.jp}
\emailAdd{kondok@faculty.chiba-u.jp}
\emailAdd{akihiro.shibata@kek.jp}
\emailAdd{skato@oyama-ct.ac.jp}

\abstract{The lattice gauge-scalar model with the scalar field in the fundamental representation of the gauge group has a single confinement-Higgs phase which is well-known as the Fradkin-Shenker-Osterwalder-Seiler analytic continuity theorem: Confinement and Higgs regions are subregions of an analytically continued single phase and there are no thermodynamics phase transitions between them.
In this talk, however, we show that we can define new type of operators which enable to separate completely the confinement phase and the Higgs phase. In fact, they are constructed in the gauge-invariant procedure by combining the original scalar field and the so-called color-direction field which is obtained by change of field variables based on the gauge-covariant decomposition of the gauge field due to Cho-Duan-Ge-Shabanov and Faddeev-Niemi.
We perform the numerical simulations for the model with SU(2) gauge group without any gauge fixing and find a new transition line which agrees with the conventional thermodynamic transition line in the weak gauge coupling and divides the confinement-Higgs phase into two separate phases, confinement and the Higgs, in the strong gauge coupling. All results are obtained in the gauge-independent way, since no gauge fixing has been imposed in the numerical simulations.
Moreover, we give a physical interpretation for the new transition from the viewpoint of the spontaneous breaking of a global symmetry.
This talk is based on the preprint \cite{IKKS23}.}

\FullConference{The 40th International Symposium on Lattice Field Theory (Lattice 2023)\\
July 31st - August 4th, 2023\\
Fermi National Accelerator Laboratory\\}

\begin{document}
\maketitle

\section{Introduction}

It is widely believed that the lattice SU(2) gauge-scalar model with a single fundamental scalar field has a single confinement-Higgs phase:
Confinement and Higgs regions are subregions of an analytically continued single phase and there are no thermodynamic phase transitions between them \cite{OsterwalderSeiler78,FradkinShenker79,BanksRabinovici79}.
However, physics to be realized in these regions are quite different despite the absence of thermodynamic transition. See e.g., \cite{Maas19} for a review.
Recently, Greensite and Matsuyama \cite{Greensite-Matsuyama18,Greensite-Matsuyama20} proposed a criterion based on the global symmetry called the \textit{custodial symmetry}.

We propose new gauge-invariant composite operators which enable to discriminate between the confinement phase and the Higgs phase in the lattice SU(2) gauge-fundamental scalar model. 
The new operators are constructed gauge-independently by combining the original fundamental scalar field and the \textit{color-direction field} which is obtained by change of field variables based on the gauge-covariant decomposition of the gauge field (CDGSFN decomposition) \cite{Cho8081,DuanGe79,Shabanov99,FaddeevNiemi9907}, see \cite{KKSS15} for a review. 
This type of operator was already introduced for investigating the phase structure of the lattice SU(2) gauge-adjoint scalar model to show the existence of the transition line which divides the confinement phase into two parts \cite{ShibataKondo23}. 

Then we investigate the phase structure of the above model by performing the numerical simulations without any gauge fixing.
As a gauge-independent result, we find a new transition line which separates the confinement-Higgs phase into two different phases, the confinement phase and the Higgs phase, in the strong gauge coupling region, in addition to reproducing the conventional thermodynamic transition line in the weak gauge coupling region.
We argue that the two regions are discriminated by the symmetric or broken realization of a global symmetry.

\section{Lattice SU(2) gauge-fundamental scalar model}

We introduce the lattice SU(2) gauge-scalar model with a single scalar field in the fundamental representation of the gauge group where the radial degrees of freedom of the scalar field is fixed. The action with the gauge coupling constant $\beta$ and the scalar coupling constant $\gamma$ is given by
\begin{align}
 S[U,\hat{\Theta}]
  = \frac{\beta}{2} \sum_{x,\mu>\nu} \re \tr \left( \mathbf{1} - U_{x,\mu} U_{x+\mu,\nu} U_{x+\nu,\mu}^{\dagger} U_{x,\nu}^{\dagger} \right)
  +\frac{\gamma}{2} \sum_{x,\mu} \re \tr \left( \mathbf{1} - {\hat{\Theta}}_x^{\dagger} U_{x,\mu} {\hat{\Theta}}_{x+\mu} \right) ,
\label{action1}
\end{align}
where $U_{x,\mu} \in \mathrm{SU(2)}$ is a (group-valued) gauge variable on a link $\langle x,\mu \rangle$, and ${\hat{\Theta}}_x \in \mathrm{SU(2)}$ is a (matrix-valued) scalar variable in the fundamental representation on a site $x$ which obeys the unit-length (or radially fixed) condition: ${\hat{\Theta}}^{\dagger}_x {\hat{\Theta}}_x = \bm{1} ={\hat{\Theta}}_x {\hat{\Theta}}^{\dagger}_x$.

This action is invariant under the local $\mathrm{SU(2)}_\mathrm{local}$ gauge transformation and the global $\widetilde{\mathrm{SU(2)}}_\mathrm{global}$ transformation. Therefore, this model has the $\mathrm{SU(2)}_\mathrm{local} \times \widetilde{\mathrm{SU(2)}}_\mathrm{global}$ symmetry:
\begin{align}
  U_{x,\mu} &\mapsto U_{x,\mu}^{\prime} = {\Omega}_x U_{x,\mu} {\Omega}_{x+\mu}^{\dagger} \, , \quad {\Omega}_x \in \mathrm{SU(2)}_\mathrm{local} , \notag\\
  \hat{\Theta}_x &\mapsto \hat{\Theta}_x^{\prime} = {\Omega}_x \hat{\Theta}_x \Gamma, \quad \Gamma \in \widetilde{\mathrm{SU(2)}}_\mathrm{global} .
\end{align}



In our investigations, the color-direction field plays the key role. This new field was introduced in the framework of change of field variables \cite{KKSS15} which is originally based on the gauge-covariant decomposition of the gauge field due to Cho-Duan-Ge-Shabanov\cite{Cho8081,DuanGe79,Shabanov99} and Faddeev-Niemi\cite{FaddeevNiemi9907}. 

The \textit{color-direction field} $\bm{n}_x$ is a local gauge-covariant site variable defined by
\begin{align}
 \bm{n}_x := n_x^A {\sigma}^A \in \mathrm{su(2)-u(1)} \quad (A=1,2,3) \, , \quad \bm{n}_x \mapsto \bm{n}_x^{\prime} = {\Omega}_x \bm{n}_x {\Omega}_x^{\dagger} \, ,
\end{align}
where ${\sigma}^A$ are the Pauli matrices. $\bm{n}_x$ has the unit length $\bm{n}_x \cdot \bm{n}_x = 1$.

For a given gauge field configuration $\{ U_{x,\mu} \}$, we determine the color-direction field configuration $\{ \bm{n}_x \}$ (as the unique configuration up to the global color rotation) by minimizing the so-called \textit{reduction functional} $F_\mathrm{red} [\bm{n} ; U]$ under the gauge transformations:
\begin{align}
 F_\mathrm{red} [\{ \bm{n}\}  ; \{ U\}]
  := \sum_{x,\mu} \frac{1}{2} \tr \left[ {\left( D_{\mu} [U] \bm{n}_x \right)}^{\dagger} \left( D_{\mu} [U] \bm{n}_x \right) \right] 
  = \sum_{x,\mu} \tr \left( \bm{1} - \bm{n}_x U_{x,\mu} \bm{n}_{x+\mu} U_{x,\mu}^{\dagger} \right) \, .
\label{red}
\end{align}
In this way, a set of color-direction field configurations $\{ \bm{n}_x \}$ is obtained as the (implicit) functional of the original link variables $\{ U_{x,\mu} \}$, which is written symbolically as 
\begin{align}
 \bm{n}^*= \underset{\bm{n}}{\text{argmin}} \, F_\mathrm{red} [\{ \bm{n}\}  ; \{ U\}].
\label{redc2}
\end{align}
This construction shows the non-local nature of the color-direction field. 


We proceed to investigate the phase structure of the model.
First, we measured the averages of the plaquette action density $P$ and the scalar action density $M$ defined by
\begin{align}
 P
  = \frac{1}{6V} \sum_{x,\mu<\nu} \tr \left( U_{x,\mu} U_{x+\mu,\nu} U_{x+\nu,\mu}^{\dagger} U_{x,\nu}^{\dagger} \right) \, , \quad
 M
  = \frac{1}{4V} \sum_{x,\mu} \tr \left( \hat{\Theta}_x^{\dagger} U_{x,\mu} \hat{\Theta}_{x+\mu} \right) \, ,
\end{align}
where $V$ is the total number of sites on the lattice.

Moreover, it is possible to define a new gauge-invariant operator $\bm{r}_x$, which is constructed from the original fundamental scalar field $\hat{\Theta}_x$ and the color-direction field $\bm{n}_x$. First we introduce a local gauge-invariant  \textit{scalar-color composite field} $\bm{r}_x$ which however transforms under the global transformation in the covariant way:
\begin{align}
 \bm{r}_x &:= \hat{\Theta}_x^{\dagger} \bm{n}_x \hat{\Theta}_x = \bm{r}_x^\dagger \, , \quad
 \bm{r}_x \mapsto \bm{r}_x^{\prime} = \Gamma^\dagger \bm{r}_x \Gamma \, .
\end{align}

Then we define the gauge-invariant \textit{scalar-color composite field density} $\bm{R}$ as the spacetime average of $\bm{r}_x$, which has the same global transformation property as $\bm{r}_x$:
\begin{align}
 \bm{R} &:= \frac{1}{V} \sum_x \bm{r}_x  = \frac{1}{V} \sum_x \hat{\Theta}_x^{\dagger} \bm{n}_x \hat{\Theta}_x = \bm{R}^\dagger \, , \quad
 \bm{R} \mapsto \bm{R}^{\prime} = \Gamma^\dagger \bm{R} \Gamma \, .
\end{align}
It should be remarked that $\bm{R}$ is not contained in the original action, in sharp contrast to the operators $P$ and $M$.
Notice that every component of the matrix $\bm{R}$ is gauge-invariant, but it is not invariant under the global transformation. Therefore, in order to show gauge-independently the spontaneous breaking of the global symmetry, we have only to measure one of the component of the matrix $\bm{R}$:
\begin{align}
 \bm{R}
  := R^A {\sigma}^A
  = \begin{pmatrix} R^3 & R^1-i R^2 \\ R^1+i R^2 & -R^3 \end{pmatrix} \in \mathrm{su(2)} \, ,
 \quad R^A = \frac{1}{2} \tr (\sigma^A \bm{R}) \quad (A=1,2,3) \, .
\end{align}

We need to take into account all the components on equal footing simultaneously to examine the spontaneous breaking of the global symmetry correctly.

From this viewpoint, we define the gauge-invariant \textit{norm} as an order parameter by
\begin{align}
 {\left\| \bm{R} \right\|}_n  := \left( \sum_{A=1}^{3}  |R^A|^n \right)^{1/n} \, ,
\end{align}
which is expected to reflect the correlation between the color-direction field $\bm{n}_x$ and the fundamental scalar field $\hat{\Theta}_x$, and detect the spontaneous breaking of the global symmetry $\widetilde{\mathrm{SU(2)}}_\mathrm{global}$.

The $n=1$ case is just the sum of all the components which is not invariant under any continuous subgroup of the global group $\widetilde{\mathrm{SU(2)}}_\mathrm{global}$, and hence can be used to show the complete spontaneous breaking of the global symmetry $\widetilde{\mathrm{SU(2)}}_\mathrm{global}$:
\begin{align}
 {\left\| \bm{R} \right\|}_1  = |R^1| + |R^2| + |R^3| \, .
\label{abr3d}
\end{align}

The $n=2$ case is equivalent to the scalar-color composite density \textit{norm} $\left\|  \bm{R} \right\|_2$ which is invariant under both the local gauge and global transformations:
\begin{align}
 {\left\| \bm{R} \right\|}_2
  = \sqrt{  {(R^1)}^2 + {(R^2)}^2 + {(R^3)}^2 }
  = \sqrt{\frac{1}{2} \tr ( \bm{R}^{\dagger} \bm{R} )} \, , \quad
 {\left\| \bm{R} \right\|}_2 &\mapsto \left\| \bm{R}^{\prime} \right\|_2 = {\left\| \bm{R} \right\|}_2 \, .
\label{abr3e}
\end{align}


To see the meaning of ${\left\| \bm{R} \right\|}_2$, we obtain the eigenvalues of $\bm{R}$  by solving the characteristic equation for the eigenvalue problem:
\begin{align}
 0 &= \det (\bm{R} - {\lambda} \bm{1})
  = (\lambda-\lambda_{+})(\lambda-\lambda_{-}) \, , \quad
 {\lambda}_{\pm} 
  = \pm \sqrt{\bm{R}^2} := \pm \sqrt{  {(R^1)}^2 + {(R^2)}^2 + {(R^3)}^2 } \, .
\label{eigenv}
\end{align}
Therefore, the scalar-color density $\bm{R}$ can be transformed into the diagonal form and the norm ${\left\| \bm{R} \right\|}_2 $ consists of two eigenvalues of the scalar-color density $\bm{R}$: $\lambda={\lambda}_{\pm} := \pm \sqrt{\bm{R}^2}$.

\section{Numerical simulations}

We performed the Monte Carlo simulations for $144$ sets of couplings $(\beta,\gamma)$ on the $8^4$ and $16^4$ lattice.
The configuration of $\{ U_{x,\mu} \}$ and $\{ \hat{\Theta}_x \}$ were updated by the pseudo heat bath method (with Kennedy-Pendleton method \cite{KennedyPendleton85} for large $\beta, \gamma$).
For a measurement with a set of couplings $(\beta,\gamma)$, we discarded first 5000 sweeps and sampled configurations per 100 sweeps and stored 100 configurations.
For each configuration $\{ U_{x,\mu} \}$, we obtained the color-direction field configuration $\{ \bm{n}_x \}$ by using the iterative method with over-relaxation to solve the reduction condition.

\begin{figure}[!t]
\centering\vspace{-5mm}
\begin{subfigure}{75mm}
  \centering\includegraphics[width=75mm]{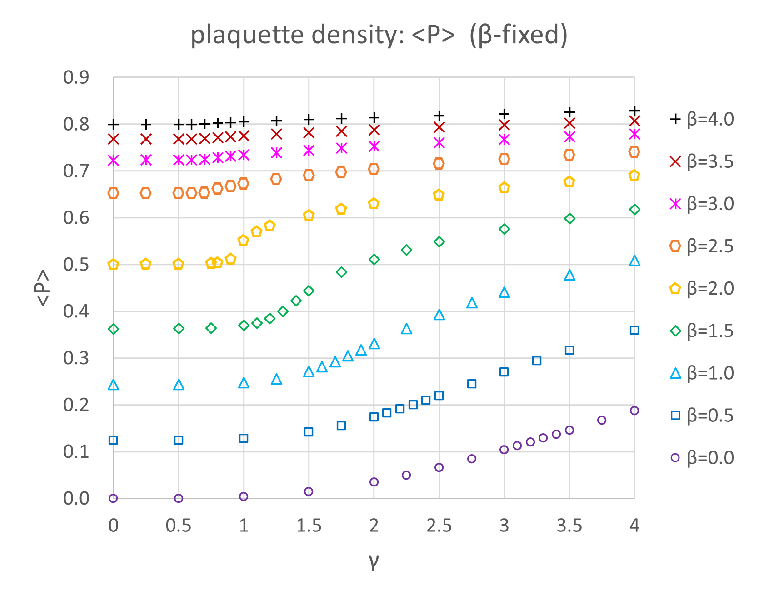}
  \label{pa1}
\end{subfigure}
\begin{subfigure}{75mm}
  \centering\includegraphics[width=75mm]{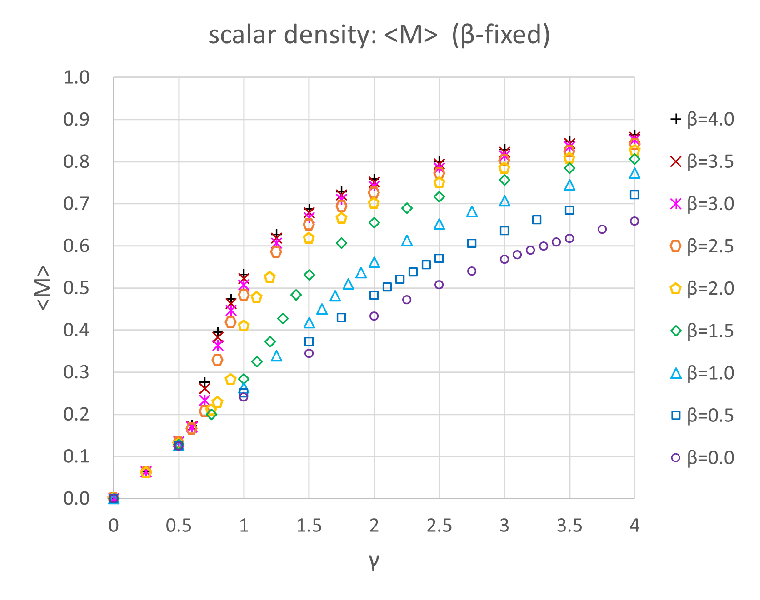}
  \label{sa1}
\end{subfigure}
\vspace{-10mm}\caption{Average of the plaquette action density $\langle P \rangle$ and the scalar action density $\langle M \rangle$ on the $16^4$ lattice: (Left) $\langle P \rangle$ vs. $\gamma$ on various $\beta = \mathrm{const.}$ lines, (Right) $\langle M \rangle$ vs. $\gamma$ on various $\beta = \mathrm{const.}$ lines.}
\label{ps1}
\end{figure}

First, we determine the transition line from the plaquette action density $\langle P \rangle$ and the scalar action density $\langle M \rangle$.
Fig.\ref{ps1} shows the measurement results of $\langle P \rangle$ and $\langle M \rangle$ in the $\beta$-$\gamma$ phase plane. The left panel is the plots of $\langle P \rangle$, while the right panel is the plots of $\langle M \rangle$ as functions of $\gamma$ on various $\beta = \mathrm{const.}$ lines. In these plots, error bars are omitted because errors are too small to be indicated.

\begin{figure}[!t]
\centering\vspace{-2.5mm}
\begin{subfigure}{75mm}
  \centering\includegraphics[width=75mm]{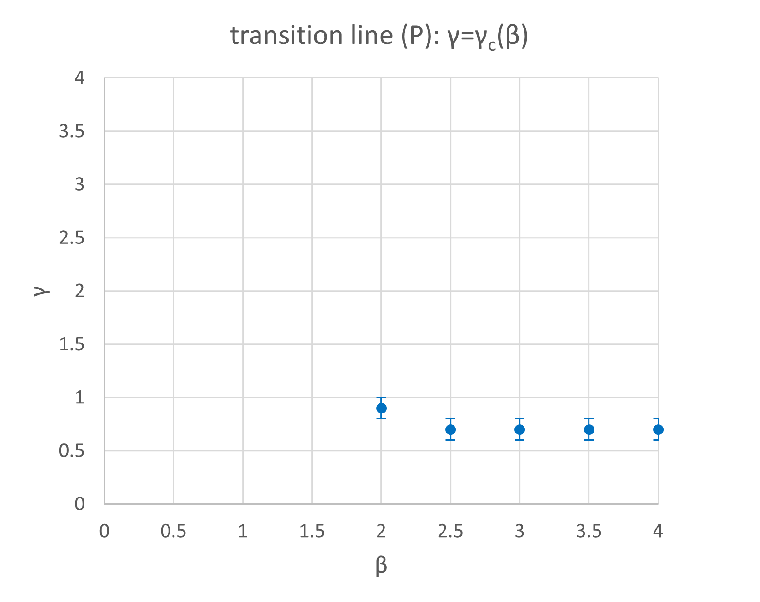}
  \label{pad}
\end{subfigure}
\begin{subfigure}{75mm}
  \centering\includegraphics[width=75mm]{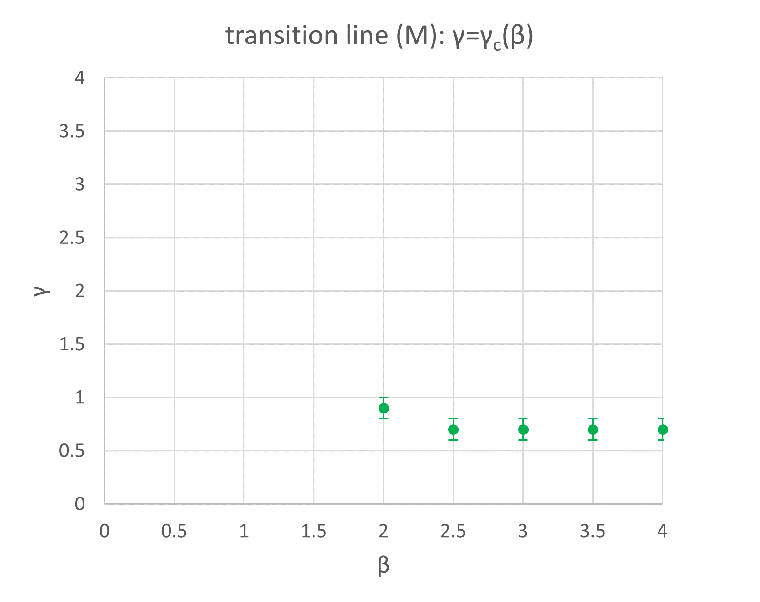}
  \label{sad}
\end{subfigure}
\vspace{-10mm}\caption{Transition lines $\gamma = {\gamma}_c (\beta)$ determined by the action densities on the $16^4$ lattice:  (Left) $P$, (Right) $M$.}
\label{psd}
\end{figure}

Fig.\ref{psd} is the transition line determined from $\langle P \rangle$ and $\langle M \rangle$, by observing gaps in these plots.
Notice that these transition lines obtained from $\langle P \rangle$ and $\langle M \rangle$ agree with each other within the errors. Then we can conclude that we reproduced gauge-independently the transition line which was obtained in the specific gauge \cite{LangRebbiVirasoro81}.

Next we determine the transition line from the scalar-color composite field density $\bm{R}$. We have proposed to measure the average $\langle {\left\| \bm{R} \right\|}_n \rangle$ to search the new transition. The global symmetry $\widetilde{\mathrm{SU(2)}}_\mathrm{global}$ is unbroken if $\langle {\left\| \bm{R} \right\|}_n \rangle \to 0$ as $V \to \infty$, while the global symmetry $\widetilde{\mathrm{SU(2)}}_\mathrm{global}$ is broken if $\langle {\left\| \bm{R} \right\|}_n \rangle \to \mathrm{const.} > 0$ as $V \to \infty$.
However, even in the unbroken phase, $\langle {\left\| \bm{R} \right\|}_n \rangle$ takes the non-zero value $\langle {\left\| \bm{R} \right\|}_n \rangle =: \langle {\left\| \bm{R}_0 \right\|}_n \rangle \neq 0$ when the lattice volume $V$ is finite.

Let $\bm{r}_{0,x}$ be the random field variable on the surface $S^2$ which has the same global transformation property as $\bm{r}_x$: $\bm{r}_{0,x} \mapsto \Gamma^\dagger \bm{r}_{0,x} \Gamma$.
Then we introduce another gauge-invariant field density $\bm{R}_0$:
\begin{align}
 \bm{R}_0 := \frac{1}{V} \sum_x \bm{r}_{0,x} = \bm{R}_0^{\dagger} \, , \quad
 \bm{R}_0 \mapsto \Gamma^\dagger \bm{R}_0 \Gamma \, .
\end{align}
We can estimate the volume dependence as $\langle {\left\| \bm{R}_0 \right\|}_n \rangle \propto \frac{1}{\sqrt{V}}$ which yields $\langle {\left\| \bm{R}_0 \right\|}_n \rangle \to 0$ as $V \to \infty
$.
In order to detect the spontaneous breaking of the global symmetry $\widetilde{\mathrm{SU(2)}}_\mathrm{global}$ in the finite volume $V$, therefore, we redefine the average of the \textit{gauge-invariant operator norm} $\langle {\left\| \bm{R} \right\|}_n \rangle^\mathrm{sub}$ by
\begin{align}
 \langle {\left\| \bm{R} \right\|}_n \rangle^\mathrm{sub} := \langle {\left\| \bm{R} \right\|}_n \rangle - \langle \left\| \bm{R}_0 \right\|_n \rangle \, .
\label{abr3c}
\end{align}
$\langle {\left\| \bm{R} \right\|}_n \rangle^\mathrm{sub}$ is the well-defined order parameter:
 $\langle {\left\| \bm{R} \right\|}_n \rangle^\mathrm{sub} =  0$ in the $\widetilde{\mathrm{SU(2)}}_\mathrm{global}$ unbroken phase, and $\langle {\left\| \bm{R} \right\|}_n \rangle^\mathrm{sub} \neq 0$ in the $\widetilde{\mathrm{SU(2)}}_\mathrm{global}$ broken phase.

\begin{figure}[!h]
\centering
\begin{subfigure}{75mm}
  \centering\includegraphics[width=75mm]{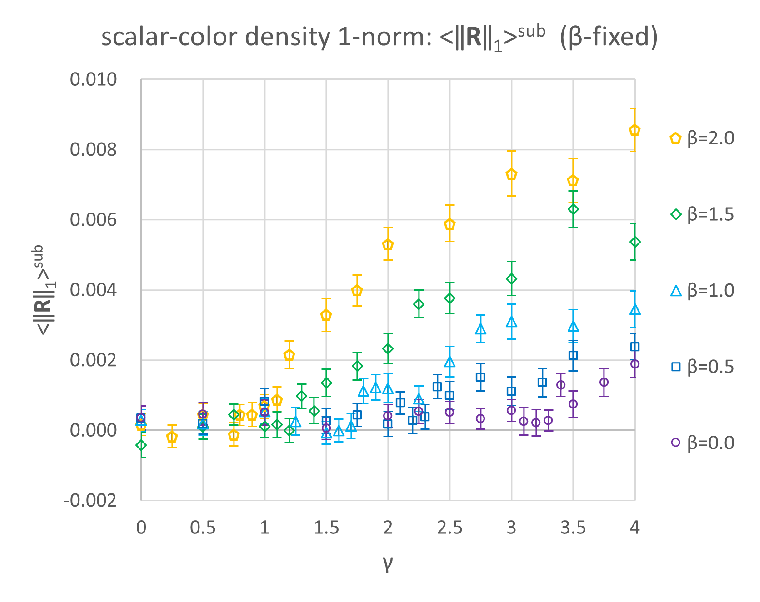}
  \label{ta1a}
\end{subfigure}
\begin{subfigure}{75mm}
  \centering\includegraphics[width=75mm]{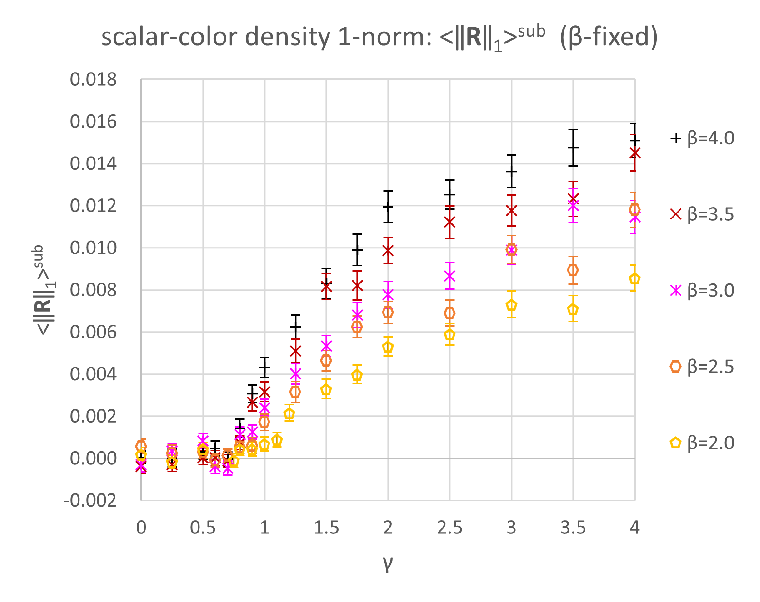}
  \label{ta1b}
\end{subfigure}\\
\begin{subfigure}{75mm}
  \centering\includegraphics[width=75mm]{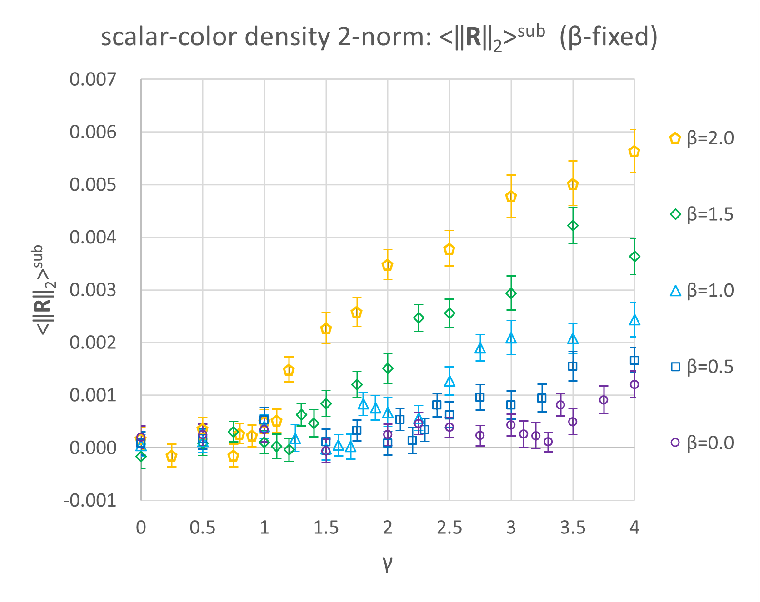}
  \label{ra1a}
\end{subfigure}
\begin{subfigure}{75mm}
  \centering\includegraphics[width=75mm]{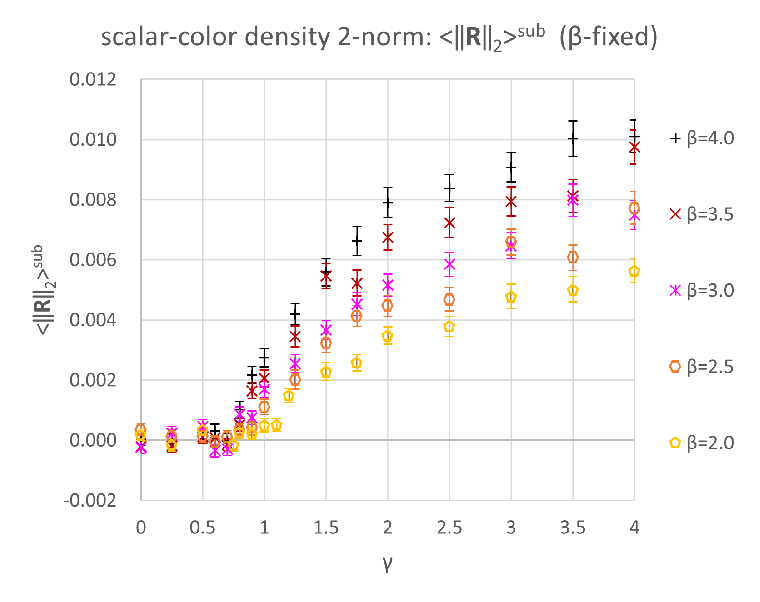}
  \label{ra1b}
\end{subfigure}
\vspace{-10mm}\caption{Average of the 1-norm $\langle {\left\| \bm{R} \right\|}_1 \rangle^\mathrm{sub}$ and 2-norm $\langle {\left\| \bm{R} \right\|}_2 \rangle^\mathrm{sub}$ of the scalar-color composite field density $\bm{R}$ on the $16^4$ lattice: (Upper) $\langle {\left\| \bm{R} \right\|}_1 \rangle^\mathrm{sub}$ vs. $\gamma$ on various $\beta = \mathrm{const.}$ lines, (Lower) $\langle {\left\| \bm{R} \right\|}_2 \rangle^\mathrm{sub}$ vs. $\gamma$ on various $\beta = \mathrm{const.}$ lines.} 
\label{tr1}
\end{figure}

\begin{figure}[!h]
\centering\vspace{-2.5mm}
\begin{subfigure}{75mm}
  \centering\includegraphics[width=75mm]{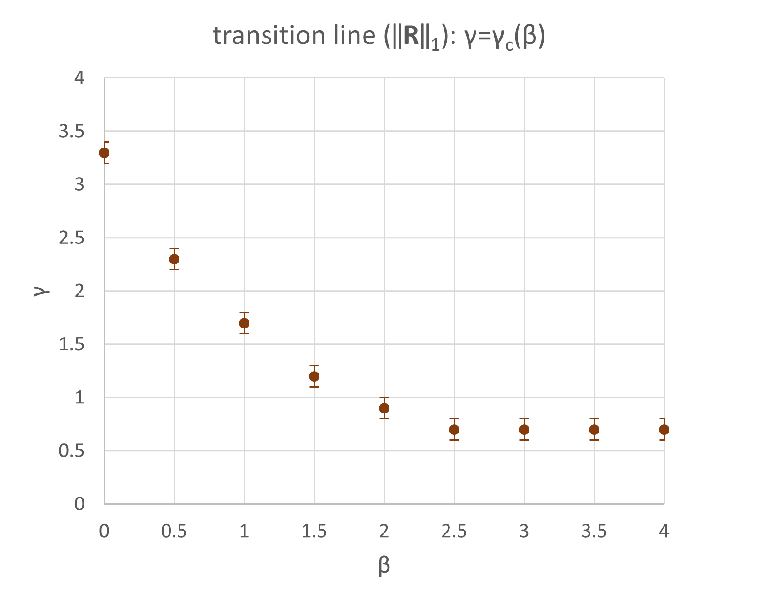}
  \label{tad}
\end{subfigure}
\begin{subfigure}{75mm}
  \centering\includegraphics[width=75mm]{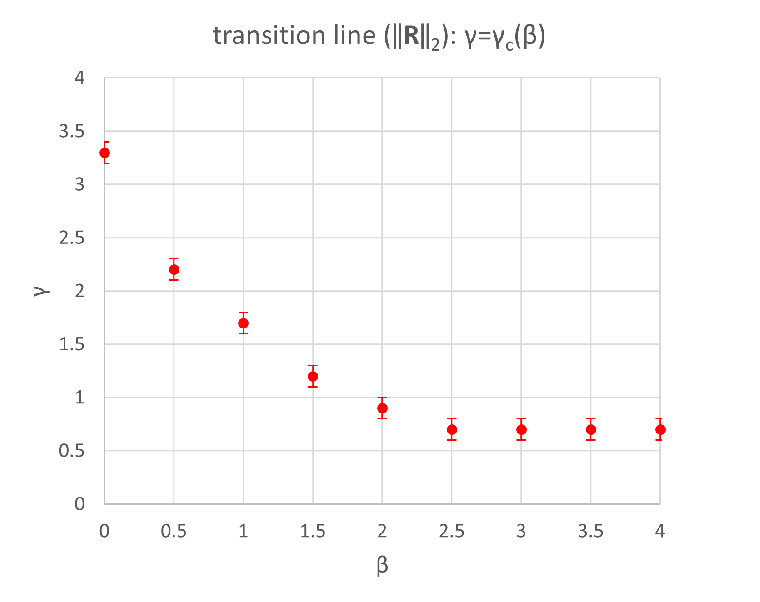}
  \label{rad}
\end{subfigure}
\vspace{-10mm}\caption{Transition lines $\gamma = {\gamma}_c (\beta)$ determined by the norms of the scalar-color composite field density on the $16^4$ lattice: (Left) ${\left\| \bm{R} \right\|}_1$, (Right) ${\left\| \bm{R} \right\|}_2$.}
\label{trd}
\end{figure}

Then we measured the redefined 1-norm $\langle {\left\| \bm{R} \right\|}_1 \rangle^\mathrm{sub}$ and 2-norm $\langle {\left\| \bm{R} \right\|}_2 \rangle^\mathrm{sub}$.
To determine the transition line, we observed the position at which the value of $\langle {\left\| \bm{R} \right\|}_n \rangle^\mathrm{sub}$ as a function of the $\beta$ changes from zero $\langle {\left\| \bm{R} \right\|}_n \rangle^\mathrm{sub} = 0$ to a non-zero value $\langle {\left\| \bm{R} \right\|}_n \rangle^\mathrm{sub} > 0$.

Fig.\ref{tr1} gives the measurement results of $\langle {\left\| \bm{R} \right\|}_1 \rangle^\mathrm{sub}$ and $\langle {\left\| \bm{R} \right\|}_2 \rangle^\mathrm{sub}$ in the $\beta$-$\gamma$ phase plane. The upper panels are the plots of $\langle {\left\| \bm{R} \right\|}_1 \rangle^\mathrm{sub}$, while the lower panels are the plots of $\langle {\left\| \bm{R} \right\|}_2 \rangle^\mathrm{sub}$ as functions of $\gamma$ on various $\beta = \mathrm{const.}$ lines.

Fig.\ref{trd} is the transition line determined from $\langle {\left\| \bm{R} \right\|}_1 \rangle^\mathrm{sub}$ and $\langle {\left\| \bm{R} \right\|}_2 \rangle^\mathrm{sub}$, by observing the results of Fig.\ref{tr1}.
It is remarkable that these new transition lines divide the single Higgs-confinement region into two separated regions: the confinement region and the Higgs region.
Notice that these transition lines obtained from $\langle {\left\| \bm{R} \right\|}_1 \rangle^\mathrm{sub}$ and $\langle {\left\| \bm{R} \right\|}_2 \rangle^\mathrm{sub}$ in the gauge-independent manner, agree with each other within the errors.
\footnote{
Incidentally, it should be mentioned that the new transition line dividing the confinement phase in the case of the adjoint scalar field has been found quite recently in \cite{ShibataKondo23}, by performing gauge-independent numerical simulations in the similar way.
}

\section{New phase structure}

\begin{figure}[!h]
\centering\vspace{-7.5mm}
\begin{subfigure}{60mm}
  \centering\includegraphics[width=60mm]{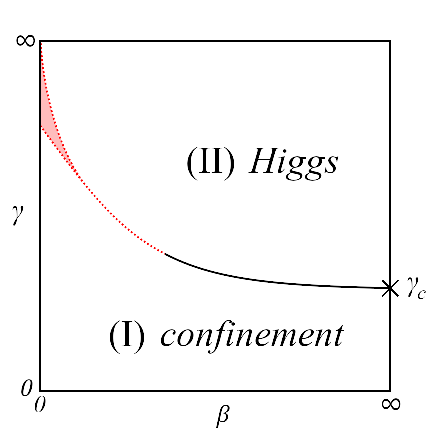}
\end{subfigure}
\vspace{3mm}
\caption{The schematic phase diagram: (I) confinement phase and (II) Higgs phase. 
The red area describes the possible locations of the new transition line due to finite volume effects.}
\label{2ph}
\end{figure}

According to our numerical simulations, the phase diagram is divided into Confinement phase (I) $\gamma<\gamma_c(\beta)$ ($\langle {\left\| \bm{R} \right\|}_2 \rangle^\mathrm{sub} = 0$) and Higgs phase (II) $\gamma>\gamma_c(\beta)$ ($\langle {\left\| \bm{R} \right\|}_2 \rangle^\mathrm{sub} \neq 0$) as shown schematically in Fig.\ref{2ph}.

First, we discuss that Confinement phase (I) and Higgs phase (II) can be respectively characterized by the absence or presence of spontaneous breaking of the global symmetry $\widetilde{\mathrm{SU(2)}}_\mathrm{global}$.

Notice that $\bm{R}$ is a Hermitian matrix. Therefore, $\bm{R}$ can be diagonalized by a unitary matrix and can be expressed using the real-valued eigenvalues $\lambda_\pm$ defined in (\ref{eigenv}) as 
\begin{align}
 \bm{R}
  = \begin{pmatrix} R^3 & R^1+iR^2 \\ R^1-iR^2 & -R^3 \end{pmatrix}
  = \Gamma_* \begin{pmatrix} \lambda_+ & 0 \\ 0 & \lambda_- \end{pmatrix} \Gamma_{*}^{ \dagger} \, , \quad
 \Gamma_* \in \widetilde{\mathrm{SU(2)}}_\mathrm{global} \, ,
\end{align}
where $\Gamma_*$ represents a certain matrix of $\widetilde{\mathrm{SU(2)}}_\mathrm{global}$ which realizes the diagonalization.
To obtain the non-vanishing average avoiding the cancellations between $\lambda_+$ and $\lambda_-$, we use only $\lambda_+>0$. 

In Higgs phase (II) ($\langle {\left\| \bm{R} \right\|}_2 \rangle^\mathrm{sub} \neq 0$), a specific rotation matrix $\Gamma_* \in \widetilde{\mathrm{SU(2)}}_\mathrm{global}$ is chosen to realize the diagonalization of the matrix $\bm{R}$ with non-zero eigenvalue $\lambda = \lambda_\pm \neq 0$.
Therefore, Higgs phase (II) is an ordered phase with the spontaneously broken global symmetry $\widetilde{\mathrm{SU(2)}}_\mathrm{global}$.

In Confinement phase (I) ($\langle {\left\| \bm{R} \right\|}_2 \rangle^\mathrm{sub} = 0$), any specific rotation matrix $\Gamma_*$ is not needed. Therefore, Confinement phase (I) is a disordered phase with the unbroken global symmetry $\widetilde{\mathrm{SU(2)}}_\mathrm{global}$.

Notice that the above argument has nothing to do with the local gauge symmetry $\mathrm{SU(2)}_\mathrm{local}$ for $\bm{R}$. The local symmetry $\mathrm{SU(2)}_\mathrm{local}$ is unbroken in both phases. Therefore, Confinement phase (I) is the phase where both the local gauge symmetry $\mathrm{SU(2)}_\mathrm{local}$ and the global symmetry $\widetilde{\mathrm{SU(2)}}_\mathrm{global}$ are unbroken, while Higgs phase (II) is the phase where the local gauge symmetry $\mathrm{SU(2)}_\mathrm{local}$ is unbroken but the global symmetry $\widetilde{\mathrm{SU(2)}}_\mathrm{global}$ is spontaneously broken.

Next, we discuss how the respective phase is characterized from the physical point of view. 

(i) 
First, we consider Confinement phase (I) $\gamma<\gamma_c(\beta)$ ($\langle {\left\| \bm{R} \right\|}_2 \rangle^\mathrm{sub} = 0$).
In this phase, confinement would occur due to vacuum condensations of appropriate topological defects, e.g., magnetic monopoles for non-Abelian gauge theory \cite{dualsuper} and the gauge fields become massive due to self-interactions among themselves.
Confinement phase (I) is regarded as a disordered phase where the color-direction field $\bm{n}_{x}$ takes the isotropic configuration in color space.
This phase is characterized by $\langle {\left\| \bm{R} \right\|}_2 \rangle^\mathrm{sub} = 0$ which means the very small correlation between the color-direction field $\bm{n}_x$ and the fundamental scalar field $\hat{\Theta}_x$.

(ii)
Next, we consider Higgs phase (II) $\gamma>\gamma_c(\beta)$ ($\langle {\left\| \bm{R} \right\|}_2 \rangle^\mathrm{sub} \neq 0$).
In this phase, the gauge fields become massive due to the absence of massless gauge mode. According to the conventional BEH mechanism, this phenomenon is understood as a consequence of the spontaneous symmetry breaking $\mathrm{SU(2)} \to \{ \bm{1} \}$. Note that the \textit{gauge-independent description of the BEH mechanism} \cite{Kondo18} provides the interpretation without introducing the spontaneous gauge symmetry breaking.
Higgs phase (II) is regarded as an ordered phase where the color-direction field $\bm{n}_x$ takes the anisotropic configuration in color space together with the fundamental scalar field $\hat{\Theta}_x$ which tends to align to an arbitrary but a specific direction.
This phase is characterized by $\langle {\left\| \bm{R} \right\|}_2 \rangle^\mathrm{sub} \neq 0$ which means the strong correlation between the color-direction field $\bm{n}_{x}$ and the fundamental scalar field $\hat{\Theta}_x$.

\section{Conclusion}

We re-examined the phase structure of the lattice SU(2) gauge-scalar model with the fundamental scalar field by introducing the new type of gauge-invariant operators. 
This model has a single confinement-Higgs phase composed of analytically continued confinement and Higgs subregions, and there are no thermodynamic phase transitions between the two regions \cite{FradkinShenker79,OsterwalderSeiler78}.
We constructed gauge-invariant composite operators composed of the fundamental scalar field and the {color-direction field} constructed from the gauge field which can be obtained from change of field variables \cite{KKSS15} based on the gauge-covariant decomposition of the gauge field due to Cho-Duan-Ge-Shabanov \cite{Cho8081,DuanGe79,Shabanov99} and Faddeev-Niemi \cite{FaddeevNiemi9907}.
We performed the gauge-fixing-free numerical simulations and found a new gauge-independent transition line which divides a single confinement-Higgs phase into the confinement phase and the Higgs phase in the strong gauge coupling region, while it reproduces the conventional thermodynamic transition line in the weak gauge coupling region. 
We provided a possible physical understanding of the resulting separated phases as a symmetric and spontaneously broken realization of a global symmetry $\widetilde{\mathrm{SU(2)}}_\mathrm{global}$. 



\clearpage
\section*{Acknowledgements}
This work was supported by JST, the establishment of university fellowships towards the creation of science technology innovation, Grant Number JPMJFS2107.
This work was also supported by Grant-in-Aid for Scientific Research, JSPS KAKENHI Grant
Number (C) No.23K03406. 
This research was supported in part by the Multidisciplinary
Cooperative Research Program in CCS, University of Tsukuba. 
The numerical simulation is supported by the High Energy Accelerator Research Organization (KEK).


\begin{thebibliography}{99}
\bibitem{IKKS23}
R. Ikeda, S. Kato, K.-I. Kondo, and A. Shibata,
Preprint CHIBA-EP-259, in preparation.
arXiv:2308.13430 [hep-lat]

\bibitem{OsterwalderSeiler78}
K. Osterwalder and E. Seiler,
Ann. Phys. {\bf 110}, 440 (1978).

\bibitem{FradkinShenker79}
E. Fradkin and S.H. Shenker,
Phys. Rev. D{\bf 19}, 3682 (1979).

\bibitem{BanksRabinovici79}
T. Banks and E. Rabinovici,
Nucl. Phys. B{\bf 1609}, 349 (1979).

\bibitem{Maas19}
A. Maas, 
Prog. Part. Nucl. Phys. {\bf 106}, 132 
 (2019).
arXiv:1712.04721 [hep-ph]

\bibitem{Greensite-Matsuyama18}
J. Greensite and K. Matsuyama, 
Phys. Rev. D{\bf 98}, 074504  (2018). 
arXiv:1805.00985 [hep-th]

\bibitem{Greensite-Matsuyama20}
J. Greensite and K. Matsuyama, 
Phys. Rev. D{\bf 101}, 054508 (2020).
arXiv:2001.03068 [hep-th]

\bibitem{Cho8081}
Y.M. Cho,
Phys. Rev. D{\bf21}, 1080 (1980);\\
Y.M. Cho,
Phys. Rev. D{\bf23}, 2415 (1981).

\bibitem{DuanGe79}
Y.S. Duan and M.L. Ge,
Sinica Sci. {\bf11}, 1072 (1979).

\bibitem{Shabanov99}
S.V. Shabanov,
Phys. Lett. B{\bf463}, 263 (1999). [hep-th/9907182]

\bibitem{FaddeevNiemi9907}
L.D. Faddeev and A.J. Niemi,
Phys. Rev. Lett. {\bf82}, 1624 (1999). [hep-th/9807069];\\
L.D. Faddeev and A.J. Niemi,
Nucl. Phys. B{\bf776}, 38 (2007). [hep-th/0608111]

\bibitem{KKSS15}
K.-I. Kondo, S. Kato, A. Shibata and T. Shinohara, Phys.
Rept. \textbf{579}, 1--226 (2015). arXiv:1409.1599 [hep-th]

\bibitem{ShibataKondo23}
A. Shibata and K.-I. Kondo,
(2023). 
arXiv:2307.15953 [hep-lat]











\bibitem{KennedyPendleton85}
A.D. Kennedy and B.J. Pendleton,
Phys. Lett. B{\bf 156}, 393 (1985).


\bibitem{LangRebbiVirasoro81}
C. B. Lang, C. Rebbi, and M. Virasoro,
Phys. Lett. B{\bf 104}, 294 (1981).

\bibitem{dualsuper}
Y. Nambu,
Phys. Rev. D\textbf{10}, 4262 (1974). \newline G. 't Hooft, in: High Energy
Physics, edited by A. Zichichi (Editorice Compositori, Bologna, 1975).
\newline S. Mandelstam,
Phys. Rept. \textbf{23}, 245 (1976).


\bibitem{Kondo18}
K.-I. Kondo,
Eur. Phys. J. C{\bf 78}, 577 (2018).
arXiv:1804.03279 [hep-th]






\end{thebibliography}
\end{document}